\documentclass[runningheads]{llncs}
\usepackage{graphicx}

\usepackage{textgreek}
\usepackage{csquotes}
\usepackage{multirow}
\usepackage{booktabs}
\usepackage{amsmath}
\usepackage{longtable}
\usepackage{perpage} 
\MakePerPage{footnote} 
\usepackage{ulem}
\usepackage[colorlinks=true, allcolors=blue]{hyperref}
\usepackage{subfiles}
\usepackage{enumitem}
\usepackage{pifont}
\usepackage{xspace}

\usepackage{xcolor}
\begin{document}

\title{Navigating the Thin Line: Examining User Behavior in Search to Detect Engagement and Backfire Effects}

%
\titlerunning{User Behavior in Search to Detect Engagement and Backfire Effects}
%
%
\author{Federico Maria Cau\orcidID{0000-0002-8261-3200} \and \\
Nava Tintarev\orcidID{0000-0003-1663-1627} 
}
\authorrunning{Cau and Tintarev}
%
%
\institute{Maastricht University, Maastricht, Netherlands \\
\email{\{federico.cau,n.tintarev\}@maastrichtuniversity.nl}
}

\maketitle              
\begin{abstract} 
Opinionated users often seek information that aligns with their preexisting beliefs while dismissing contradictory evidence due to confirmation bias. This conduct hinders their ability to consider alternative stances when searching the web. Despite this, few studies have analyzed how the diversification of search results on disputed topics influences the search behavior of highly opinionated users.
To this end, we present a preregistered user study ($n$ = 257) investigating whether different levels (low and high) of bias metrics and search results presentation (with or without AI-predicted stances labels) can affect the stance diversity consumption and search behavior of opinionated users on three debated topics (i.e., atheism, intellectual property rights, and school uniforms). 
Our results show that exposing participants to (counter- attitudinally) biased search results increases their consumption of attitude-opposing content, but we also found that bias was associated with a trend toward overall fewer interactions within the search page. 
We also found that 19\% of users interacted with queries and search pages but did not select any search results. When we removed these participants in a post-hoc analysis, we found that stance labels increased the diversity of stances consumed by users, particularly when the search results were biased. Our findings highlight the need for future research to explore distinct search scenario settings to gain insight into opinionated users' behavior.


\keywords{confirmation bias \and search behavior \and bias metrics.}
\end{abstract}

\section{Introduction}
Recent studies reveal that certain biases in the way users interact with web search engines can lead to unintended consequences. On the one hand, we have users' cognitive biases such as \textit{confirmation bias}, where people tend to consume information that aligns with their preexisting beliefs and ignore contradictory information when searching the web \cite{azzopardi_search_behaviour,nickerson_conf_bias}. 
On the other hand, search results from popular search engines can be \textit{biased} toward particular viewpoints (e.g., the degree of whether a document is against, neutral, or in favor of a specific topic), which can strongly influence user opinions~\cite{draws2023viewpointdiversity,lien_filter_bubbles2022,puschmann2019BubbleAssessingDiversity}.
Previous literature has shown that these phenomena may lead users to consume biased content, causing the \textit{search engine manipulation effect} (SEME). For instance, this effect occurs when users alter their attitudes based on the viewpoints expressed in highly-ranked search results~\cite{allam2014ImpactSearchEngine,draws2021ThisNotWhat,epstein2015SearchEngineManipulation,pogacar2017_thepositiveandnegative}.
As a consequence, it is necessary to implement strategies that reduce biased content presented and consumed by users, bringing the need to increase the \textit{viewpoint diversity} in search results \cite{gao2020CreatingFairerRanking,mattis_nudging_framework_2021,mcdonald2022,eval_fairness_argument_retrieval_2021}.

Many attempts have already been made to mitigate the SEME \cite{draws_fairness1,mattis_nudging_framework_2021,rieger2021ThisItemMight,Xu2021_HowDoUsersOpinions}. Nevertheless, most of these studies focused on mildly opinionated users, while content consumption and behaviors of users with strong opinions (i.e., \textit{opinionated}) are still under-explored. In contrast, a recent study by Wu et al. \cite{zhangyi2023_exp} found that predicted stance labels and their explanations increase content diversity consumption compared to plain search results when considering opinionated users, although the benefits of such explanations over stance labels remain unclear.
Furthermore, it is uncertain if their results on the effect of stance labels on content diversity consumption were affected by: \textbf{i)} the use of fixed, hand-made templates for generating biased/balanced search results lists without proper validation from state-of-the-art bias metrics \cite{draws2023viewpointdiversity,draws_fairness1}, or \textbf{ii)} the abundance of neutral search results at the top of the list in the balanced condition, or \textbf{iii)} the lack of investigation for potential users' behaviors triggered by their exposure to predicted stance labels \cite{azzopardi_search_behaviour,epstein2017SuppressingSearchEngine,rieger2021ThisItemMight}.

In this work, we address these gaps and investigate the effects of different viewpoint bias metric levels and predicted stance labels of search results on opinionated users' clicking diversity and search strategies. 

We produce balanced and biased search result lists by defining two levels of \textit{polarity} (nDPB) and \textit{stance} (nDSB) normalized discounted viewpoint bias metrics \cite{draws2023viewpointdiversity} for both opposing and supporting pre-study attitude of a user towards a debated topic (four conditions in total; see Table \ref{tab:conditions_scores}). 
We assign each user to a random SERP (same set of 40 search results) of the 40 differently ranked SERPs for a specific condition. Hence, each SERP consists of the same 40 search results but ranked differently based on the bias metric level and the user pre-study attitude. Whenever a user issues a query, we randomly show a different SERP ranking (out of the 40 SERPs) that still fulfills the study conditions the user was initially assigned (see Section \ref{ranked_list_serp}).
We focus on two research questions:

\vspace{.5em}

\noindent \textbf{RQ1.} How do SERPs informed by different levels of bias metrics and AI-generated stance labels impact users’ diversity of search results they click on?

\noindent \textbf{RQ2.} How do SERPs informed by different levels of bias metrics and AI-generated stance labels impact users’ search behaviors?

\vspace{.5em}

To address our research questions, we conducted a preregistered, online user study ($N=257$; Section \ref{sec:participants}) simulating a low-stakes and open-ended search scenario. We asked opinionated participants to inform themselves regarding a debated topic with the help of search results containing two different bias levels and with or without informative stance labels. 
Our study shows that \textit{participants exposed to biased search results tended to (i) consume more attitude-opposing content} and (ii) \textit{interact less within the search page}. Interestingly, we found that 19\% of users abandoned \cite{query_abandonment2020} the web search, probably triggered by a backfire effect \cite{backfire_2020}.
When we excluded these participants in a post-hoc analysis, we discovered that stance labels increased the diversity of stances consumed by users, especially when the search results were biased. 
Although we did not find evidence of specific search behavior traits among different bias metrics and search results presentations, we discuss in Section \ref{sec:discussion} how our results call for further research to promote responsible online opinion formation. The study preregistration, materials, and data used in the paper are available at \url{https://osf.io/ph8nv/}.
\section{Related Work and Hypotheses}

\subsection{Viewpoint Diversity and Ranking Algorithms}
\label{sec:viewpoint_div}
\textit{Viewpoints} \cite{viewpoints_arguments2019,draws2022ComprehensiveViewpointRepresentationsa,viewpoints_arguments2020,kucuk2021StanceDetectionSurvey} are opinions relative to debated topics or claims \cite{draws2022ComprehensiveViewpointRepresentationsa}, and can be represented in ranked lists of search results by a label assigned to each document, which can be expressed in different degrees (e.g., binary or multi-categorical \cite{chenSeeingThingsDifferent2019,draws2020HelpingUsersDiscoverb,draws2023viewpointdiversity,geziciEvaluationMetricsMeasuring2021,pogacar2017_thepositiveandnegative,yomtovPromotingCivilDiscourse2014}).
A lack of viewpoint diversity can be seen as a lens to \textit{bias} in search engines \cite{gao2020CreatingFairerRanking}, which calls for measuring and promoting \textit{exposure diversity} \cite{mattis_nudging_framework_2021} of search results to prevent SEME for users when searching the web.
Earlier work investigated how to mitigate search results biases through various approaches. One approach is to generate fair (or less biased) rankings by diversifying search results for a pre-defined disadvantaged viewpoint category \cite{mcdonald2022}, measuring the degree of bias in search engines by studying the correlations between statistical parity (e.g., equal exposure of different viewpoint categories) and search results diversity, fairness, and relevance \cite{gao2020CreatingFairerRanking,eval_fairness_argument_retrieval_2021}, and establishing normative diversity definitions to evaluate news recommendation systems \cite{Vrijenhoek_radio}.

An exception is a recent paper by Wu et al. \cite{zhangyi2023_exp} which investigated how different levels of bias in search results influenced the behavior (diversity of viewpoints selected) of opinionated users.
 
However, they found no evidence of differences between the two bias conditions. 
We hypothesize that a possible reason for this outcome may lie in the balanced and biased fixed templates they used (i.e., always having five stronger stances or neutral results at the top but without the use of viewpoint bias metrics), while we propose 40 different SERP combinations per bias level for two viewpoint bias metrics (nDPB and nDSB) \cite{draws2023viewpointdiversity}. For our first hypothesis, we expect users to consume less diverse content when exposed to a SERP with a high bias.

\vspace{.5em}
\noindent \textit{Hypothesis 1a} (\textbf{$\text{H}_{\text{1a}}$}). Users exposed to SERPs with \textit{high} bias metrics will interact with less diverse results than users exposed to SERPs with \textit{low} bias metrics.

\subsection{Content Diversity Consumption and Search Results Labels}
\label{sec:content_consumption}
Beyond biases in the presentation of the search results during online searches, another significant challenge comes from the user side. This phenomenon is \textit{confirmation bias}, where users tend to favor and consume information that aligns with their existing beliefs \cite{rieger2021ThisItemMight}. 
Specifically, as in Section \ref{sec:viewpoint_div} we discussed strategies to address \textit{exposure diversity} (i.e., the share of supply diversity that a user is exposed to \cite{mattis_nudging_framework_2021}), here we are interested in investigating how we can achieve \textit{consumption diversity} (i.e., the overall diversity in content that users actively engage with \cite{mattis_nudging_framework_2021}). 
Recent research \cite{mattis_nudging_framework_2021}, proposed a framework where they distinguish five potential diversity nudges (i.e., system architecture design choices that predictably alter users' behavior) that can be used to increase the content consumption diversity. Specifically, they suggested how interface design interventions (e.g., presentation nudges) can make users' content consumption more transparent. For example, incorporating informative labels can help reduce confirmation bias in search results \cite{epstein2017SuppressingSearchEngine,rieger2021ThisItemMight}.
Furthermore, stance detection models \cite{sanh2019distilbert} can identify the stance or viewpoint (e.g., against, neutral, supporting) of search results and have been proven to be effective in increasing the diversity of users' content consumption and help them navigate online debates \cite{zhangyi2023_exp}. While some studies exist \cite{mapping_exposure_diversity2022,is_this_a_click_towards_diversity2021,benefits_diverse_news2022} which investigated strategies to increase users' content diversity consumption, the effects of these interventions on specifically \textit{opinionated users} are still under-explored. 

In this work, we aim to investigate the effectiveness of predicted stance labels (i.e., against, neutral, and supporting) 
in mitigating confirmation bias on search results ranked according to two bias metrics levels. Moreover, if stance labels effectively reduce confirmation bias, we can anticipate an interaction effect between the bias condition in search results and the interface design (i.e., search results with and without stance labels).
The hypotheses follow the work of Wu et al. \cite{zhangyi2023_exp}:

\vspace{.5em}
\noindent \textit{Hypothesis 1b} (\textbf{$\text{H}_{\text{1b}}$}). Users exposed to search results with \textit{stance labels} interact with more diverse content than users who are exposed to regular search results.

\vspace{.5em}
\noindent \textit{Hypothesis 1c} (\textbf{$\text{H}_{\text{1c}}$}). Users exposed to search results with \textit{stance labels} are less susceptible to the effect of stance biases in search results on clicking diversity.

\subsection{Users and Search Behavior}
Earlier works found evidence that users' search behaviors are affected by factors like search results' viewpoint biases, user knowledge of the topic, and warning labels for biased search results
\cite{epstein2017SuppressingSearchEngine,suppanut2019_AnalyzingTheEffects,rieger2021ThisItemMight,Suzuki2021_AnalysisOfRelationship,Suzuki2021CharacterizingTI,Xu2021_HowDoUsersOpinions}. 
For example, the authors of \cite{suppanut2019_AnalyzingTheEffects} found that users spent more effort searching by \textit{issuing more queries} when they encountered documents that were inconsistent with their prior beliefs. 
Xu et al. \cite{Xu2021_HowDoUsersOpinions} found that (i) users' opinions seem to be easily affected by search results when they browse the web search purposelessly, and (ii) users who strongly support a topic \textit{issued more clicks} and \textit{spent more time} on the search results. 
Further, Epstein et al. \cite{epstein2017SuppressingSearchEngine} findings suggest that alerts for biased search results reduced the SEME effect and induced users to \textit{issue more clicks}, \textit{spend more time searching} and explore \textit{lower-ranked search results}.
Nevertheless, these works focused on mildly opinionated users, while there is still a lack of studies investigating opinionated users' behavior. 
We expect that users' behavior will vary based on the interaction between the SERP bias and display. Specifically, considering high bias metrics and search results incorporating stance labels, we hypothesize users will spot that search results are biased in their attitude-opposing direction when first inspecting the search engine. Consequently, users will not readily select the top attitude-opposing results (while trying to find information to confirm their beliefs) but instead i) issue more queries, ii) make fewer clicks, iii) visit more pages to check whether they are also biased, iv) have a higher click depth, and v) spend a longer time searching.

\vspace{.5em}
\noindent \textit{Hypothesis 2} (\textbf{$\text{H}_{\text{2a-e}}$}). Users exposed to SERPs informed by \textit{high} bias metrics and incorporating \textit{stance labels} will issue more queries (\textbf{$\text{H}_{\text{2a}}$}), fewer clicks (\textbf{$\text{H}_{\text{2b}}$}), visit more pages (\textbf{$\text{H}_{\text{2c}}$}), click on deeper search results (\textbf{$\text{H}_{\text{2d}}$}), and spend more time (\textbf{$\text{H}_{\text{2e}}$}) 
in the search session than users who are exposed to SERPs informed by \textit{low} bias metrics with regular search results. 
\section{Experimental Setup}
This section describes the user study we conducted to investigate the effect of SERP bias, display, and their interaction on users' behavior. 

\subsection{Materials}
\label{sec:materials}

\subsubsection{Dataset.}
To compute viewpoint bias metrics and predict stances for our study, we considered a public data set containing 1453 search results of queries related to three different debated topics (i.e., \textit{atheism}, \textit{intellectual property rights}, and \textit{school uniforms})~\cite{draws2023viewpointdiversity} and used for stance detection in previous works~\cite{draws2023explainable,zhangyi2023_exp}. Each search result in the dataset has been annotated in an expert annotation procedure and assigned to a \textit{viewpoint label}, which indicates its position in the debate surrounding the topic it refers to on a seven-point Likert scale ranging from ``strongly opposing'' (-3) to ``strongly supporting'' (3). We mapped the viewpoints labels into three categories: 
\textit{against} (-3,-2,-1), \textit{neutral} (0), and \textit{supporting} (1,2,3) (i.e., suitable ternary classification for our stance detection model).
Next, we followed the same procedure as Wu et al. \cite{zhangyi2023_exp} to process the search results for the stance detection model, obtaining 1125 final instances.

\subsubsection{Viewpoint Diversity Metric.}
To define the SERP bias levels, we considered two metrics from the \textit{normalized discounted viewpoint bias} (nDVB) \cite{draws2023viewpointdiversity}, which consider viewpoints on a seven-point Likert scale (-3,+3) to compute biases.  
The first, \textit{normalized discounted polarity bias} (nDPB), evaluates the extent to which opposing and supporting viewpoints are balanced and ranges from -1 to 1. (more extreme values indicate greater bias, values closer to 0 indicate neutrality).
The second metric, \textit{normalized discounted stance bias} (nDSB), measures how much the distribution of stances deviates from being evenly spread across the seven different stance categories, and ranges from 0 (desired scenario of stance diversity) to 1 (maximal stance bias). 

\begin{table}[]
    \centering
    \scriptsize
    \setlength{\tabcolsep}{5pt}
    \caption{Summary statistics of our four sets of 40 SERPs for polarity (nDPB) and stance (nDSB) bias metrics considering \textit{low-low} and \textit{high-high} levels for the \textit{supporting} and \textit{opposing} bias directions. 
    All the values are identical among topics.}
    \begin{tabular}{llllcccc}
        \toprule
        \# & \textbf{Bias Metric} & \textbf{Level} & \textbf{Bias Direction} & \textbf{Min} & \textbf{Max} & \textbf{Mean} & \textbf{Std.} \\
        \midrule
        \multirow{2}{2em}{I} & Polarity  & low & supporting & 0.05 & 0.12 & 0.09 & 0.02 \\ 
        & Stance  & low & supporting & 0.23 & 0.33 & 0.28 & 0.03 \\ 
        \multirow{2}{2em}{II} & Polarity  & high & supporting & 0.29 & 0.44 & 0.35 & 0.05 \\ 
        & Stance & high & supporting & 0.36 & 0.44 & 0.38 & 0.02 \\ 
        \multirow{2}{2em}{III} & Polarity  & low & opposing & -0.05 & -0.12 & -0.09 & 0.02 \\ 
        & Stance  & low & opposing & 0.23 & 0.33 & 0.28 & 0.03 \\ 
        \multirow{2}{2em}{IV} & Polarity & high & opposing & -0.29 & -0.44 & -0.35 & 0.05 \\ 
        & Stance & high & opposing & 0.36 & 0.44 & 0.38 & 0.02 \\ 
        \bottomrule
    \end{tabular}
    \label{tab:conditions_scores}
\end{table}

\subsubsection{Ranked List of Search Results.}
\label{ranked_list_serp}
To create the search result ranking conditions for the two levels of bias, we randomly selected 40 AI-correctly predicted search results\footnote{Since the number of AI-correct predictions constrained us in the selection of search results, we found that 40 search results (ten results per four pages) were a suitable number to compute low and high bias metrics and mimic a realistic search scenario.} per topic from our test and validation data considering their original seven-point viewpoint. We picked a balanced sample of six strong (-3,+3), ten moderate (-2,+2), eight mild (-1,+1), and sixteen neutral (0) results to enable the creation of viewpoint-biased and viewpoint-diverse search result rankings to simulate a realistic scenario.
We decided to have two levels of low and high bias metrics for the following reasons: i) the number of correctly AI-predicted search results is not sufficient to create a neat distinction between low-high and high-low levels for nDPB and nDSB bias metrics, and ii) we wanted to create a scenario that mirrors the \textit{balanced} and \textit{biased} templates described in Wu et al. \cite{zhangyi2023_exp} work but without having the same ranking order for each condition so that search results can still change if a user enters a new query, proposing a different ranking that still fulfils the bias metric levels. 
Next, we start from baseline rankings for low and high bias metrics levels and run a Grid Search to generate over 1000 combinations of our 40 selected search results. Then, we picked 40 sets of SERPs for low and high bias levels for each topic closest to the minimum or maximum values we found in the Grid Search or baseline rankings (see Table \ref{tab:conditions_scores}).

\begin{figure}
      \centering
      \includegraphics[width=0.8\textwidth]{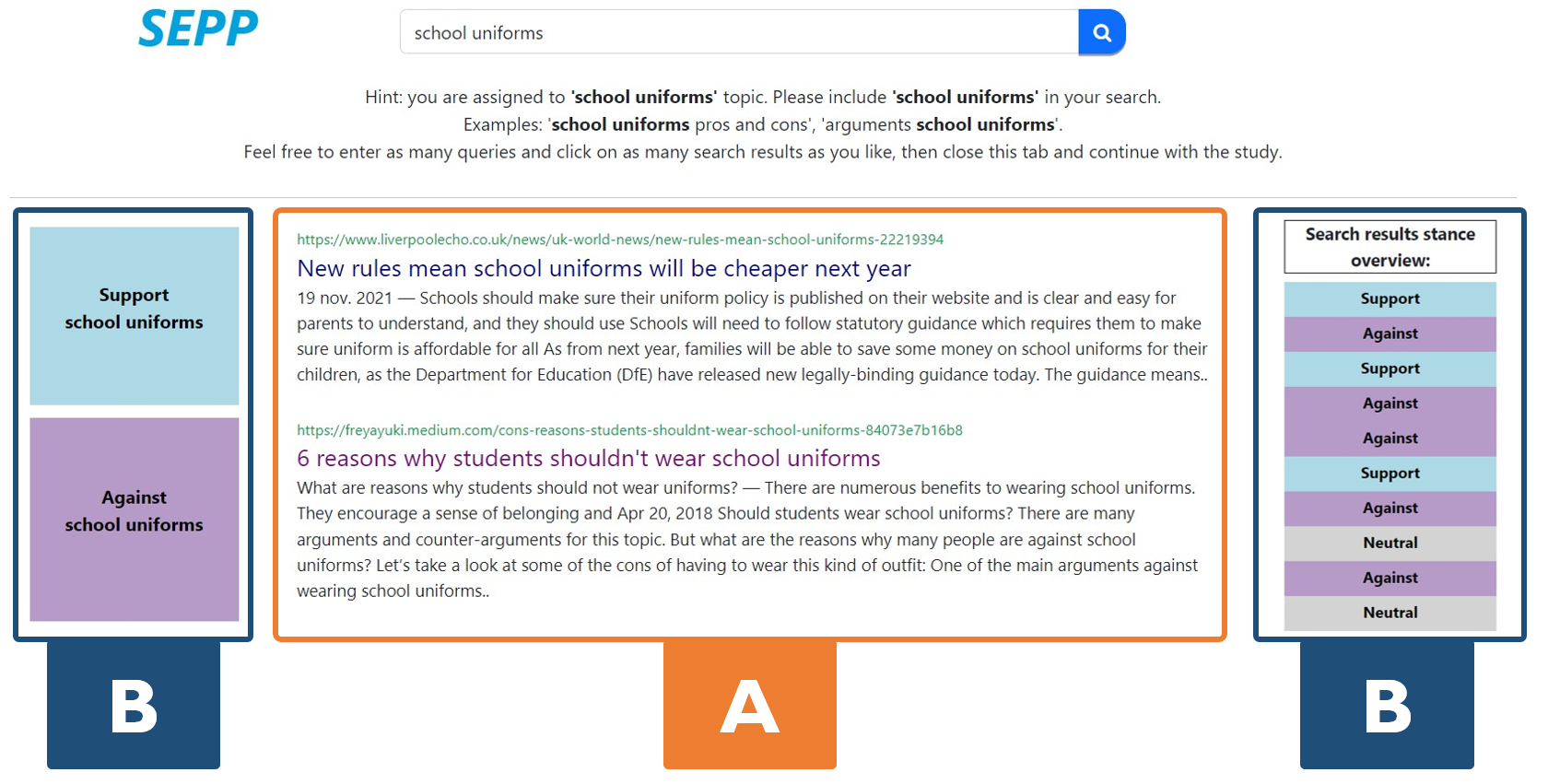}
    \caption{SERP display conditions: \textbf{A)} search results only, and \textbf{B)} search results in A) are accompanied by AI-predicted stance labels (left) and stance overview (right).}
    \label{fig:serp_display}
\end{figure}

\subsection{Variables}
\label{sec:variables}
Our study followed a between-subjects design, considering three categorical \textit{independent variables}: \textbf{SERP bias} (two levels: low or high nDPB and nDSB), \textbf{SERP display} (two levels: without stance labels or with stance labels; Fig. \ref{fig:serp_display}), and \textbf{topic} (three levels: atheism, intellectual property rights, and school uniforms).
We investigated their effect on the following numerical \textit{dependent variables}: (i) \textbf{Clicking diversity} is measured with the Shannon index to quantify the diversity of users' clicks \cite{zhangyi2023_exp}. The minimum value of the Shannon index is 0, which indicates that there is no diversity and only one viewpoint was clicked on. (ii) \textbf{Search behavior} is a set of five users' interactions, consisting of the following dependent variables: \textbf{queries}, \textbf{clicks}, \textbf{pages}, \textbf{click depth}, and \textbf{search session time}.

In addition to the above variables, we collected the following \textbf{descriptive} and \textbf{exploratory measurements} 
to describe our sample and for exploratory analyses: \textit{gender, age group, level of education}, \textit{topic knowledge} (seven-point Likert scale ranging from ``not knowledgeable at all'' to ``extremely knowledgeable''), and \textit{post search experience} (open text).\footnote{\textit{``Please shortly describe your experience with the web search engine. Did you look for specific information, and if yes, how did you try to find it? Did you think the web search helped you build a more informed opinion on [topic]? If yes/no, why?''}}

\subsection{Procedure}
\label{sec:procedure}

\paragraph{\textbf{Step 1.}} 
After agreeing to an informed consent, participants stated their gender, age group, and level of education. We then asked participants for their attitudes concerning each debated topic (including one attention check where we specifically instructed participants on what option to select from a Likert scale) via three statements on seven-point Likert scales ranging from \textit{strongly against} (-3) to \textit{strongly supporting} (3). 
Then, we assigned participants to one of the three debated topics on which they stated a strong attitude\footnote{If participants have no strong attitude on any topic, they exit the study (fully paid). }, 
and asked them to imagine the following scenario \cite{zhangyi2023_exp}: \textit{You and your friend were having dinner together. Your friend is very passionate about a debated topic and couldn't help sharing his views and ideas with you. After the dinner, you decide to further inform yourself on the topic by conducting a web search.}

\paragraph{\textbf{Step 2.}}
We introduced participants to the task and randomly assigned them\footnote{We balanced participation across topics, bias, and display experimental conditions.} to one SERP bias condition\footnote{We randomly assign participants to one of the 40 search results combinations with an opposite bias direction based on their pre-stance attitude (i.e., users who strongly support a topic are assigned to the opposing bias direction and vice versa).} and one SERP display condition (see Section \ref{sec:variables}).
We asked participants to click on a link that leads them to our search platform.  
Here, we asked participants to issue as many queries and click on as many search results as they like, as long as those queries include their assigned topic term (e.g., \enquote{arguments school uniforms} for the topic \textit{school uniforms}).
Every time they entered a new query, they received a different set of the 40 SERPs based on their assigned topic and bias metric conditions  
(see Section \ref{sec:materials}). 
Once participants were done searching, we asked them to return to the survey page and advance to the next step.\footnote{To encourage interactions with the web search engine, participants could only advance to the next step of the study after one minute. There was no maximum search time to simulate a realistic scenario.}

\paragraph{\textbf{Step 3.}} 
We again measured participants' attitudes on their assigned topic and asked them to provide textual feedback on their search experience. We included another attention check - where the answer was explicitly reported in the question text - to filter out low-quality data.

\subsection{Planned Sample Size and Statistical Analysis}
\label{sec:participants}
Before data collection, we had computed a required sample size of $247$ participants using the software \textit{G*Power} \cite{faul2009statistical} for a between-subjects ANOVA, specifying a medium effect size (Cohen's $f$ = 0.25), a desired power of $0.8$, and four groups. 
We recruited participants aged 18 or older with high English proficiency from \textit{Prolific} \footnote{\url{https://prolific.co}}. The task  was hosted on \textit{Qualtrics} \footnote{\url{https://www.qualtrics.com/}}. Participants received £1.4 as a reward for the study\footnote{The study has been approved by the Ethics Committee of Maastricht University.} (i.e., £13.3/hour given the median task completion time of 6:05 minutes). Prolific automatically timed out participants after 44 minutes. We further excluded participants from the analysis if they did not have a strong stance on any of the three topics, and failed at least one attention check.
We investigated \textbf{$\text{H}_{\text{1a-c}}$} by conducting a \textbf{between-subjects ANOVA} with \textit{clicking diversity} as the dependent variable and studying the main and interaction effects of \textit{SERP bias} and \textit{SERP display}, using \textit{topic} as an additional independent variable (confounding factor).
For \textbf{$\text{H}_{\text{2a-e}}$}, we conducted five \textbf{between-subjects ANOVA}  considering  \textit{search behaviors} (see Section \ref{sec:variables}) as dependent variables and studying interaction effects of \textit{SERP bias} and \textit{SERP display}, while adding \textit{topic} to the analysis to control for its potential role as a confounding factor.
Since we assessed eight ANOVA tests as part of this study, we applied a Bonferroni correction to our significance threshold, reducing it to $\frac{0.05}{8}=0.0063$. We will consider as significant only the \textit{p}-values that are below this specific threshold.

\section{Results}
We recruited 350 paid participants, excluding them from the analysis if they did not have a strong opinion on any of the three topics (33), and failed at least one attention check (60). We thus considered 257 participants for the analysis.

\subsection{Descriptive Analysis}
Participants' age distribution was 66\% in the 25-44 group, 8\% in the 18-25 group, while 26\% were above 44 y.o., consisting of 51\% females, 48\% males, and 1\% other. 
Participants educational background show that 30\% had a high school diploma, 44\% had a bachelor's degree, 18\% had a master’s degree, and 8\% had a Doctorate or professional degree. 
Most of the participants (49\%) had little knowledge about their assigned topic, 21\% had almost no knowledge, and 30\% were very knowledgeable.
Furthermore, the topic, bias, and display conditions distribution were approximately equally balanced among participants. 
Participants’ pre-search stances are mostly distributed towards supporting the corresponding topic: atheism (26\% vs 8\%), intellectual property rights (33\% vs $<$1\%), and school uniforms (22\% vs 10\%).  
Interestingly, 19\% (49/257) concluded their search without clicking on any search results. We will better discuss this finding in Section \ref{sec:posthoc}.

\begin{figure}
      \centering
      \includegraphics[width=0.8\textwidth]{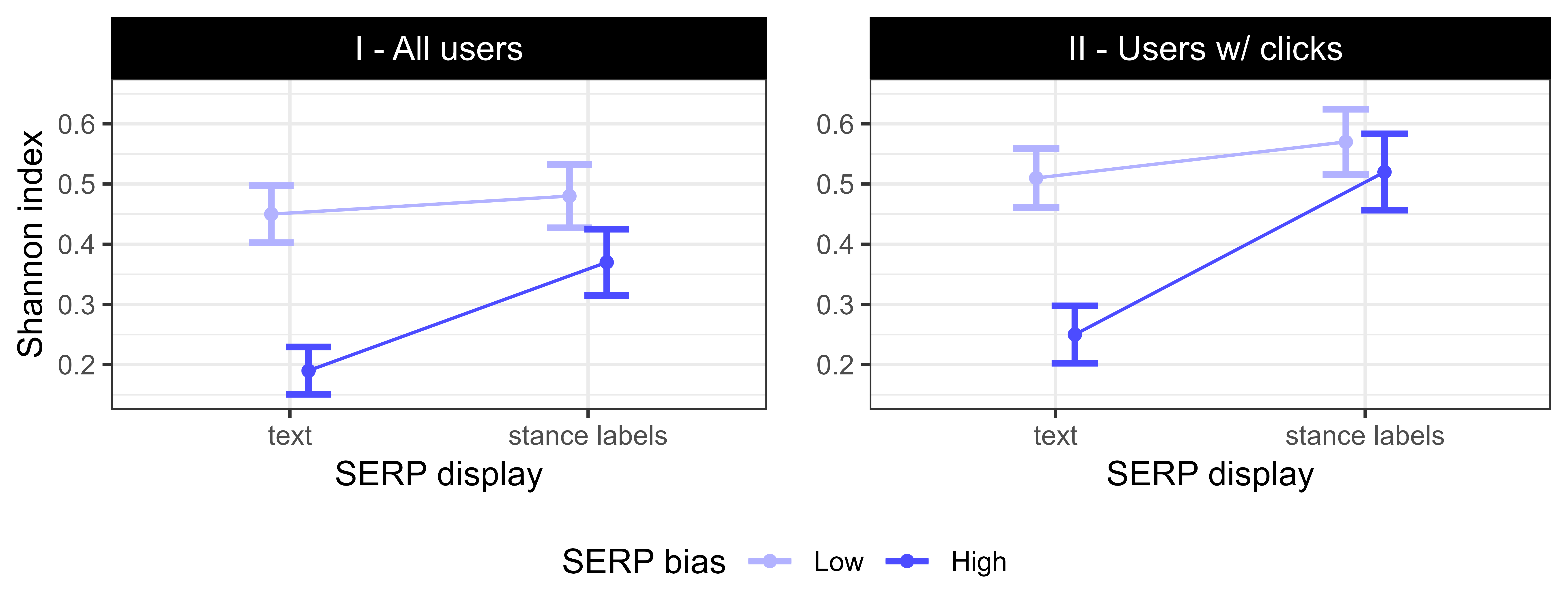}
    \caption{Shannon index split by SERP display and bias conditions considering \textbf{I)} all users (257), and \textbf{II)} users who issued clicks in the search platform (208). The error bars are computed using the Standard Error (SE) on the Shannon index.}
    \label{fig:anovas}
\end{figure}

\subsection{Hypotheses Tests}
\label{main_hyp}
For \textbf{$\text{H}_{\text{1a}}$}, we examined the clicking diversity of users who viewed low and high SERP bias levels. The ANOVA analysis highlights a significant effect for the bias level factor (F = 13.726, $p$ $<$.001, Cohen's $f$ = 0.23), hence, we \textit{reject the null hypothesis} for $\text{H}_{\text{1a}}$, concluding that there is a difference in the clicking diversity of content between users who were exposed to SERPs with low (M = 0.47, sd = 0.41) and high (M = 0.28, sd = 0.39) bias levels (see Figure \ref{fig:anovas} I). To investigate \textbf{$\text{H}_{\text{1b}}$}, we checked the clicking diversity of participants who saw search results with and without stance labels. The ANOVA analysis shows that the SERP display factor is not significant (F = 4.964, $p$ = .026, Cohen's $f$ = 0.14) and is above our predefined significance threshold (.006). In this case, we \textit{fail to reject the null hypothesis} for $\text{H}_{\text{1b}}$, concluding that there is no difference in the clicking diversity of content between users who were exposed to search results with (M = 0.43, sd = 0.44) and without (M = 0.32, sd = 0.37) stance labels. 
  To study \textbf{$\text{H}_{\text{1c}}$}, we looked at the interaction between the SERP bias and SERP display. The ANOVA analysis resulted in a non-significant interaction (F = 1.859, $p$ = .173, Cohen's $f$ = 0.09), thus we \textit{fail to reject the null hypothesis} for $\text{H}_{\text{1c}}$. 
For the ANOVA analyses concerning \textbf{$\text{H}_{\text{2}}$} we also did not find significant effects on any of the five hypotheses we tested. Hence, we \textit{fail to reject the null hypothesis} concluding that there are no differences in terms of \textit{search behaviors} (see Section \ref{sec:variables}) w.r.t. to bias levels or labels.

\begin{figure}
      \centering
    \includegraphics[width=\textwidth]{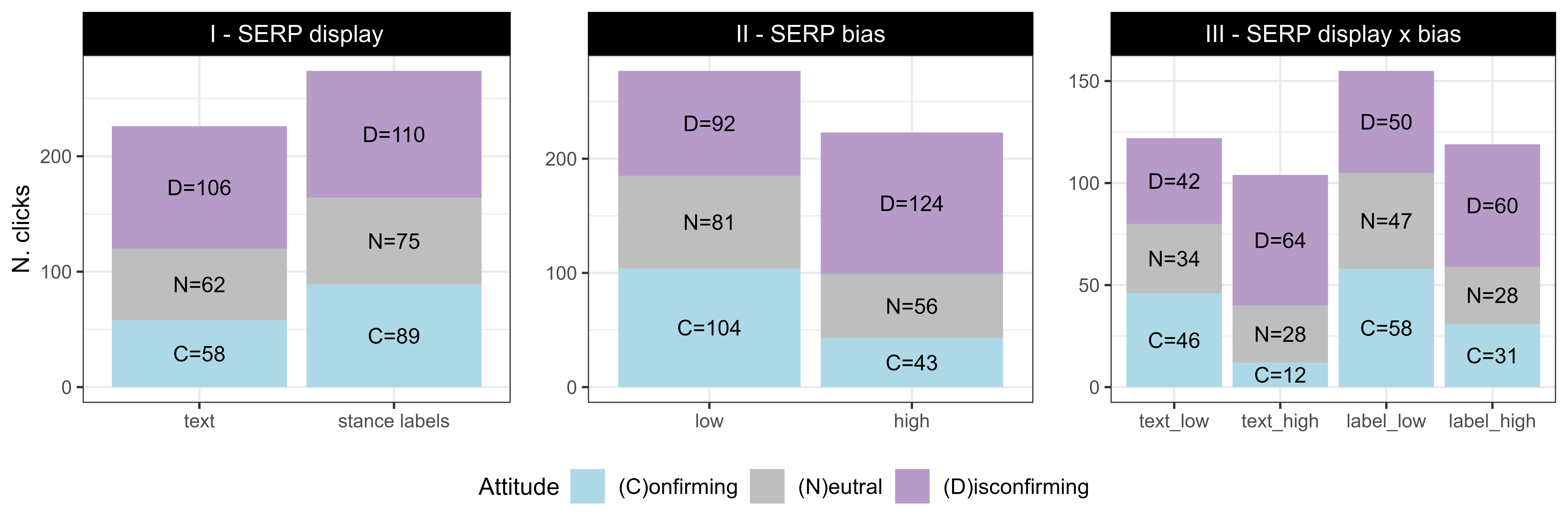}
    \caption{Users' click distribution considering different attitudes: confirming (inline with their pre-stance), neutral, and disconfirming (against their pre-stance). Conditions are grouped by \textbf{I)} SERP display, \textbf{II)} SERP bias, and \textbf{III)} SERP display and bias.}
    \label{fig:click_distr}
\end{figure}

\subsection{Post Hoc and Exploratory Analyses}
\label{sec:posthoc}
To better understand why some of the participants (49) did not click on any search results, we first discuss their behavior considering the issued queries, visited pages, and search time. Participants issued more queries (M = 2.75, sd = 3.55) and visited more pages (M = 3.04, sd = 3.50) no matter the bias level in the search results only condition. Overall, these participants had a lower search time (M = 145.52, sd = 54.60) compared to those who clicked on search results. 
Furthermore, 65\% of these users were assigned to a high level of SERP bias (37\% plain search results and 28\% to stance labels), so they perceived that search results were biased with attitude-opposing viewpoints, thus triggering a potential backfire effect that led to an abandonment \cite{query_abandonment2020} in the search task. To get more insights into their abandonment, we performed a qualitative analysis of the textual feedback provided by these participants. Some of those who were exposed to biased search results recognized the presence of bias (e.g., \enquote{\textit{...I don't think the search result helped much of my own opinion, seems lots of results are labelled as 'Against' already, which could influence the people's opinion before they read the details}}).
Additionally, we briefly discuss the trends in search behavior considering the SERP bias significant effect for all the participants (257). Search results with a high bias lead users to interact with less diverse results, but they also lead to slightly fewer interactions (i.e., participants issued fewer clicks, queries, visited fewer pages, and spent less time searching).

Next, we analyzed the clicking diversity of users who engaged more (i.e., clicked on search results) with the web search by re-examining all the hypotheses (see Figure \ref{fig:anovas} II) and comparing their behavior with users who abandoned the web search (49). 
As in Section \ref{main_hyp}, we found a significant effect (F = 8.423, $p$ $<$ .006, Cohen's $f$ = 0.20) for different levels of SERP bias (\textbf{$\text{H}_{\text{1a}}$}) and no significant interaction (F = 3.587, $p$ = .059, Cohen's $f$ = 0.13) between the SERP display and SERP bias (\textbf{$\text{H}_{\text{1c}}$}). Instead, we found a significant effect (F = 7.892, $p$ $<$ .006, Cohen's $f$ = 0.20) for different levels of SERP display (\textbf{$\text{H}_{\text{1b}}$}).
As before, we did not find significant effects on any behavior measurement (\textbf{$\text{H}_{\text{2a-e}}$}).  
The trends according to bias and display conditions are as follows. Participants exposed to SERPs with a high bias issued fewer clicks, queries, and pages visited. For the SERP display, participants exposed to stance labels issued more clicks, fewer queries, visited fewer pages, had higher click depth, and spent more time searching.

To investigate the users' clicking diversity difference on the SERP bias and display conditions, we explored the distribution across the three stances (see Figure \ref{fig:click_distr}). 
For the SERP display condition, Figure \ref{fig:click_distr} I shows a shift towards balance in attitude distribution in the stance labels condition, with an increase in the attitude-confirming results and less for the neutral ones (thus increasing the clicking diversity).
For the SERP bias condition, Figure \ref{fig:click_distr} II shows that viewpoints are somewhat balanced (slightly high confirming attitude) when the bias is low. Instead, a high bias shifts most clicks towards the \textit{disconfirming} attitude at the cost of the number of clicks.
Furthermore, Figure \ref{fig:click_distr} III reveals an increase in the number of clicks between the search results without labels and with labels when the SERP bias is low. This effect is also present with a high SERP bias in a reduced way. 
\section{Discussion}
\label{sec:discussion}
In this study, we investigated the effects of stance labels and different levels of bias metrics on opinionated users' clicking diversity and search behaviors. We conducted a between-subject study on three debated topics (i.e., \textit{atheism}, \textit{intellectual property rights}, and \textit{school uniforms}) where participants take part in an open-ended and low-stakes scenario (i.e. inform themselves on a specific topic by conducting a web search). For \textbf{RQ1}, we found that participants exposed to biased search results proportionally consume more attitude-opposing content (i.e., less diverse search results).
Furthermore, a post hoc analysis considering users who clicked on search results reveals that they consumed more diverse content in the presence of stance labels than plain search results (in line with Wu et al.  \cite{zhangyi2023_exp}). For \textbf{RQ2}, we did not find evidence of search behavior differences among our SERP bias and display conditions. However, biased search results highlighted a trend of fewer interactions within the search page (e.g., fewer clicks and less time spent searching). Furthermore, a posthoc analysis shows that users who abandoned the web search without clicking any results issued more queries, visited more pages, and spent less time searching. 

We can identify two main reasons why we did not find any significant differences in search behavior in our study. We simulated a realistic scenario and did not put any limit on the search time session and the number of interactions in general. Therefore, users could naturally explore the web search without any particular restrictions concerning search behavior. However, dealing with opinionated users considering no task restrictions and exposing them to biased search results led to a web search abandonment \cite{query_abandonment2020} for 19\% of participants. 
Second, the potential backfire effect that triggered users to abandon the web search may hint at a typical conduct present in System 1 conceptual thinking \cite{azzopardi_search_behaviour}, where people's biases affect the way they perceive and process new information, especially if the information is counter-intuitive, conflicting or induce uncertainty \cite{biases_1974}.

\subsection{Implications}
Firstly, our findings suggest that it is crucial to examine the behaviors of users who abandon their search. Our post-hoc analysis indicated that 19\% of participants abandoned the web search without interacting with any search result while issuing other search behaviors. These findings may suggest a backfire effect \cite{backfire_2020}, leading users to reject attitude-opposing content and quickly end their search. 
Second, users who interacted with search results and were exposed to stance labels consumed more diverse content, especially with biased search results. These results align with previous research on bias mitigation using informative labels \cite{epstein2017SuppressingSearchEngine,rieger2021ThisItemMight,zhangyi2023_exp}, making them a valuable tool for increasing content diversity consumption. 
These results bring the need to inspect additional factors on strongly opinionated users, like their characteristics (e.g., intellectual humility \cite{alisa_int_humility2023}), 
their perception of search results such as relevance \cite{Maxwell2019_TheImpactOfResult} or credibility \cite{suppanut2019_AnalyzingTheEffects}, and expose them to different types of tasks (high-stakes \cite{draws2021ThisNotWhat} or purposeful \cite{Xu2021_HowDoUsersOpinions}).

\subsection{Limitations and Future Work}
Our study has the following limitations. 
First, we used a ternary stance detection model to classify search results. So, users will not realize the nuances of the search results they are seeing, which are on a seven-point Likert scale. Furthermore, we did not inform users that the AI predictions on search results stances were always correct. Hence, users were unaware of the correctness of the AI-detected stances, and there is a chance they may have thought the labels were potentially wrong. Future work could investigate how to predict more diversified viewpoint representations and the effects of presenting wrong AI predictions for the detected search results stances.
Second, we decided to use the Shannon index to measure clicking diversity since we expected fewer than six clicks for each user, as suggested by previous work on users' behavior \cite{suppanut2019_AnalyzingTheEffects,rieger2021ThisItemMight,Suzuki2021_AnalysisOfRelationship}. Further studies are needed to explore other methodologies to measure users' clicking diversity based on search results viewpoints (e.g., fairness metrics \cite{awrf_fairness2019,fair_fairness2017}). Third, we performed a Grid Search to find reasonable search result pages for low and high bias metrics 
\cite{draws2023viewpointdiversity}, based on the AI correctly predicted search results and our user study conditions. Future research should investigate other methodologies to compute more fine-grained boundaries on different bias metrics and datasets.

\section{Conclusion}
In this paper, we studied the effects of stance labels and bias metrics on users' clicking diversity and search behavior.
We found that opinionated users increased their consumption of attitude-opposing content when exposed to biased search results, with a trend towards fewer interactions within the search page. Furthermore, 19\% of users who did not interact with search results spent less time searching while issuing more queries and visiting more pages. Instead, participants who issued clicks significantly increased their content diversity when exposed to stance labels, particularly when the search results were biased. 
As highlighted in our exploratory and qualitative analyses, a potential motivation for our results may lie in the open-ended task formulation, and the need to inspect additional participants' traits.
Follow-up studies need to explore whether different ranking bias metrics and stance label representations influence content diversity consumption, along with constructing different types of open and closed tasks to better investigate the search behaviors of opinionated users.

%
%
%
 \bibliographystyle{splncs04}
 \bibliography{my-bib}
\end{document}